\documentclass[12pt,twoside]{article}   
\usepackage[super,sort,comma]{natbib}

\usepackage{amsfonts}

\usepackage{fancyhdr}		

\usepackage[section]{placeins}   %

\usepackage{graphicx}

\makeatletter \renewcommand\@biblabel[1]{$^{#1}$} \makeatother
\newcommand{\cen}[1]{\begin{center} #1 \end{center}}

\usepackage{hyperref}
\hypersetup{ colorlinks,
    citecolor=blue,
    filecolor=blue,
    linkcolor=blue,
    urlcolor=blue
}

\usepackage{xcolor}

\definecolor{gray}{rgb}{0.6,0.6,0.6}
\definecolor{red}{rgb}{0.85,0,0}
\definecolor{green}{rgb}{0,0.85,0}
\definecolor{blue}{rgb}{0,0,0.85}
\definecolor{beige}{rgb}{0.92,0.87,0.78}
\usepackage[all]{hypcap}    

\begin{document}
\cen{\sf {\Large {\bfseries Probabilistic Self-learning Framework for Low-dose CT Denoising} \\  
\vspace*{10mm}
Ti~Bai$^1$, Dan~Nguyen$^1$, Biling~Wang$^1$ and~Steve~Jiang$^1$} \\
$^1$Medical Artificial Intelligence and Automation (MAIA) Laboratory, Department of Radiation Oncology, University of Texas Southwestern Medical Centre, Dallas, Texas 75239
}

\pagenumbering{roman}
\setcounter{page}{1}
\pagestyle{plain}
Corresponding Author: Steve.Jiang@UTSouthwestern.edu \\

\begin{abstract}
\noindent {\bf Purpose:} Despite the indispensable role of X-ray computed tomography (CT) in diagnostic medicine, the associated harmful ionizing radiation dose is a major concern, as it may cause genetic diseases and cancer. Decreasing patients’ exposure can reduce the radiation dose and hence the related risks, but it would inevitably induce higher quantum noise. Supervised deep learning techniques have been used to train deep neural networks for denoising low-dose CT (LDCT) images, but the success of such strategies requires massive sets of pixel-level paired LDCT and normal-dose CT (NDCT) images, which are rarely available in real clinical practice. Our purpose is to mitigate the data scarcity problem for deep learning-based LDCT denoising.\\
{\bf Methods:} To solve this problem, we devised a shift-invariant property-based neural network that uses only the LDCT images to characterize both the inherent pixel correlations and the noise distribution, shaping into our probabilistic self-learning (PSL) framework. The AAPM Low-dose CT Challenge dataset was used to train the network. Both simulated datasets and real dataset were employed to test the denoising performance as well as the model generalizability. The performance was compared to a conventional method (total variation (TV)-based), a popular self-learning method (noise2void (N2V)), and a well-known unsupervised learning method (CycleGAN) by using both qualitative visual inspection and quantitative metrics including peak signal-noise-ratio (PSNR), structural similarity index (SSIM) and contrast-to-noise-ratio (CNR). The standard deviations (STD) of selected flat regions were also calculated for comparison. \\
{\bf Results:} The PSL method can improve the averaged PSNR/SSIM values from 27.61/0.5939 (LDCT) to 30.50/0.6797. By contrast, the averaged PSNR/SSIM values were 31.49/0.7284 (TV), 29.43/0.6699 (N2V), and 29.79/0.6992 (CycleGAN). The averaged STDs of selected flat regions were calculated to be 132.3 HU (LDCT), 25.77 HU (TV), 19.95 HU (N2V), 75.06 HU (CycleGAN), 60.62 HU (PSL) and 57.28 HU (NDCT). As for the low-contrast lesion detectability quantification, the CNR were calculated to be 0.202 (LDCT), 0.356 (TV), 0.372 (N2V), 0.383 (CycleGAN), 0.399 (PSL), and 0.359 (NDCT). By visual inspection, we observed that the proposed PSL method can deliver a noise-suppressed and detail-preserved image, while the TV-based method would lead to the blocky artifact, the N2V method would produce over-smoothed structures and CT value biased effect, and the CycleGAN method would generate slightly noisy results with inaccurate CT values. We also verified the generalizability of the PSL method, which exhibited superior denoising performance among various testing datasets with different data distribution shifts. \\
{\bf Conclusions:} A deep learning-based convolutional neural network can be trained without paired datasets. Qualitatively visual inspection showed the proposed PSL method can achieve superior denoising performance than all the competitors, despite that the employed quantitative metrics in terms of PSNR, SSIM and CNR did not always show consistently better values. \\

\end{abstract}

\setlength{\baselineskip}{0.7cm}      

\pagenumbering{arabic}
\setcounter{page}{1}
\pagestyle{fancy}

\section{Introduction}
\label{sec: introduction}
X-ray computed tomography (CT) is widely used in medicine to visualize patients’ anatomy. However, it is well-known that X-rays are harmful to the human body, as ionizing radiation may cause genetic and cancerous diseases. Consequently, the \textit{as low as reasonably achievable} (ALARA) principle should be followed in radiation-related clinical practice. However, lowering the X-ray exposure to reduce the dose would inevitably induce higher quantum noise, thus yielding lower quality or even useless CT images. Therefore, a denoising algorithm is required to enhance the quality of low-dose CT (LDCT) images.

To tackle this problem, a tremendous amount of research  has been conducted, which can be classified into three different categories: pre-reconstruction projection denoising \cite{RN386,RN385}, model-based iterative reconstruction \cite{RN379,RN382,RN363,RN360,RN378,RN383,RN359,RN361} and post-reconstruction denoising \cite{RN388,RN389}. As we understand them, most of the existing algorithms can be analyzed, intrinsically speaking, from a Bayesian estimation viewpoint. In other words, they are trying to maximize a posterior by taking full advantage of the well-explored noise model of the measurements and also exploiting the potential correlations among pixels to regulate the denoised result by devising certain prior terms. For instance, the block-matching 3D filter (BM3D) \citep{RN365} assumes that there exist image patches with similar spatial structures, which can be stacked together for collaborative filtering. The total variation (TV) minimization-based methods suppose that the transformed gradient coefficients should follow a Laplacian distribution \citep{RN363,RN368}. The dictionary learning-based sparse coding techniques suggest that the image patches can be sparsely represented by a given dictionary that is pre-trained from an external high-quality image library \citep{RN360,RN359}. It should be noted that these conventional optimization-based denoising methods can be applied directly on any given single noisy image. This feature benefits from the unique driving force of these methods, i.e., the pixel correlations, which are mathematically modeled by human experts based on a comprehensive statistical analysis of clean images.

Despite their elegant theoretical design, these conventional methods are outperformed nowadays by popular data-driven deep learning (DL) techniques \citep{RN78}. Specifically, convolutional neural network (CNN)-based DL techniques, which can automatically extract strong hierarchical features from images \citep{RN138}, have achieved great success and demonstrated superior performance in various image-related tasks, such as image classification \citep{RN83,RN158}, object detection \citep{RN120}, segmentation \citep{RN87,RN326}, and super resolution \citep{RN13}. Denoising can also be regarded as an image-to-image translation task from the perspective of the computer vision community. By first extracting strong features from the original noisy image,  then employing a predictor, a DL-based denoiser can translate  the noisy image to the associated clean image \citep{RN213,RN145,RN212}.

Inspired by its unprecedented success in natural image denoising, researchers  have also used CNN to address the LDCT denoising problem. To compensate for the spatial information loss due to the down-sampling convolutional operation, a residual encoder-decoder architecture \citep{RN358} was proposed that directly connects the low-level fine features with the high-level features to facilitate better denoising performance. Shan \textit{et~at.} \citep{RN49} devised a modularized adaptive processing neural network with repeating denoising modules, which enables clinical-task–specific denoising. In addition to improvements to the network architecture, different losses \citep{RN355} have been introduced to better characterize the distance between the denoised LDCT and the normal-dose CT (NDCT), which serves as the training target. For example, Yang \textit{et~at.} \citep{RN357} developed a generative adversarial network training framework equipped with Wasserstein distance and perceptual losses, which has produced promising results. Tang \textit{et~al.} \citep{tang2019unpaired} proposed a cycle-consistent loss based unsupervised learning method for noise reduction in LDCT. All of these CNN-based denoisers have shown encouraging performance when compared with conventional methods.

However, most of the state-of-the-art CNN-based medical image denoisers employ supervised learning. It is well accepted that the availability of a rich amount of annotated training datasets is essential to supervised learning. Therefore, achieving a high denoising performance requires a massive number of paired datasets (LDCT and NDCT) so the CNN can learn an optimal mapping function between the two. Unfortunately, it is very difficult, if not impossible, to collect a large number of real patient cases for training CNN-based denoisers for multiple practical reasons. First, paired LDCT and NDCT scans from the same subjects are rarely available in routine clinical practice. Second, when researchers collect paired LDCT and NDCT scans for the same patients for research purposes, the datasets are often very small because of the radiation risk to the patients. Third, for LDCT and NDCT pairs that have been collected, the correspondence between two images of the same patient rarely has pixel-level accuracy because of patient and organ motion between the two image acquisitions. This kind of mismatch error is hard to remove through image registration and would be inevitably propagated and amplified during the DL training stage, because denoising is a dense prediction task that requires pixel-wise correspondence between input-target pairs. Noise simulation may somewhat alleviate the data scarcity challenge. However, the distribution of the noise in the CT image domain is quite complicated since it is correlated with many factors such as exposure settings, reconstruction algorithms/hyperparameters, pixel size, and slice thickness. One cannot exhaust all the possible settings for all these potential impact factors during simulation. Consequently, simulated noise may have distribution difference with the realistic noise from the CT scanner, and hence would induce model generalizability problem.
Therefore, the supervised learning-based CNN denoising algorithms suffer from the lack of perfectly paired images.

In contrast, the conventional optimization methods do not rely on paired images, and their performance does not depend on the size of datasets. The success of these pixel correlation-driven conventional methods implies that we may be able to build a medical image denoiser based on modern DL techniques that uses only noisy images if it can automatically capture and utilize the inherent correlations among pixels, which would eliminate the need for well-paired images for training. Indeed, in the field of natural image processing, to leverage CNN’s strong learning ability, several techniques have been developed to train a denoiser without ground truth images as the supervised signal. For example, Krull \textit{et~at.} \citep{RN213} proposed a NOISE2VOID (N2V) training strategy, which tries to predict any specific pixel from the surrounding pixels. To avoid the model collapsing into identity mapping, the authors adopted a random masking strategy that deliberately excludes the pixels to be predicted from the model input. The model’s performance is sub-optimal because of this exclusion. Moreover, the training efficiency of this random masking strategy is low. To alleviate these problems, Laine \textit{et~at.} \citep{RN145} proposed a unique network architecture with a built-in blind-spot mechanism. Consequently, all the predicted pixels can provide gradient information for updating weights and thereby boost the computational efficiency. The noise model can also be incorporated into the training stage to explicitly utilize the information from the input noisy image. Nonetheless, despite its higher convergence efficiency in terms of number of iterations, this method requires four times the computational burden of the N2V method in each iteration to construct the blind-spot architecture. The driving force of these deep learning-based methods is their exploitation of the inherent correlations among pixels, rather than the ground truth clean target image. Accordingly, they are usually called self-learning methods, because the input noisy image itself provides the supervised signal.

The limited availability of perfectly paired LDCT/NDCT images for supervised learning makes the self-learning framework very attractive for building a more practical LDCT denoiser. To the best of our knowledge, little effort has been devoted to this topic. In this paper, we introduce a self-learning based LDCT denoising algorithm. Specifically, we propose a new neural network architecture that automatically excavates the pixel correlations based on the inherent spatial shift-invariant properties of the image. In addition to the learned internal spatial correlations, hand-crafted prior information is also incorporated for more aggressive constraint. The entire training pipeline is well-described in the Bayesian estimation viewpoint so the informative noisy measurements can be used to produce the final denoised results, which then shape our probabilistic self-learning (PSL) framework for low-dose CT denoising.

\section{Methods and Materials}
\subsection{Mathematical formulation}

We  first formulate the LDCT denoising problem mathematically from the image translation point of view. Basically, this task translates a noisy image $Y\in \mathbb{R}^{M\times N}$ into its clean counterpart $X\in \mathbb{R}^{M\times N}$, where $M$ and $N$ denote the rows and columns, respectively. Without loss of generality, the associated noise $\epsilon \in \mathbb{R}^{M\times N}$ is assumed to be additive, such that
\begin{equation}
\label{eqn:degradation model}
Y=X+\epsilon.
\end{equation}

This problem can be solved by training a CNN $\Phi$ with parameters $W$, such that the output of  $\Phi_{W}(Y)$ is as close to $X$ as possible, given T clean-noisy image pairs $(X_t, Y_t)$, $t\in \{0,1,\cdots T-1\}$, where $t$ indexes the training sample. This is a typical supervised learning setup. If the mean squared error (MSE) loss is used, the associated cost function can be expressed as:
\begin{equation}
\label{eqn:supervised learning}
    \bar{W}=\mathrm{arg}\min_{W} \sum_{t=0}^{T-1}||\Phi_W(Y_t) - X_t||_2^{2}.
\end{equation}

As shown in equation (\ref{eqn:supervised learning}), the driving force of supervised learning originates from the strong supervised signal $X_t$, which however is nontrivial to obtain for the LDCT denoising task elaborated above. By contrast, self-learning algorithms are driven by the inherent correlations among pixels, assuming that a set of pixels $Y^E$ can be predicted by using the information from the rest of the pixels $Y^C$ in the image, where $E$ and $C$ represent the 2D coordinate sets of the pixels that satisfy $C \cup E = \{(m, n)|m\in\{0, 1,\cdots, M-1\}, n\in\{0, 1,\cdots, N-1\}\} $, $C \cap E = \emptyset$. As a result, the self-learning framework offers the possibility of training a denoiser by using only the noisy images without the corresponding clean images.

More formally, if we assume that the noisy images in the training dataset are independent, then the self-learning training process can be conducted by maximizing the following conditional probability:
\begin{equation}
\label{eqn: general self learning}
    \max\prod_{t=0}^{T-1}P(Y_t^E|Y_t^C),
\end{equation}
which can be solved by minimizing the following equivalent cost function if an independently identically distributed noise model is assumed:
\begin{equation}
\label{eqn: least square self learning}
    \bar{W}=\mathrm{arg}\min_{W} \sum_{t=0}^{T-1}||\Phi_W(Y_t^C) - Y_t^E||_2^{2}.
\end{equation}

We can consider the predicted result $\Phi_W(Y^C)$ as an unbiased denoised estimation of $Y^E$ by making the following two reasonable assumptions: 1) the noise is not intrinsically predictable, and 2) the noise on the X-ray CT image has zero mean value. One can feed different pixel sets $Y^C$ into the trained network, such that the predicted results of $Y^E$ can be composed into a complete image, thereby producing the final denoised result.

The  above training method for the self-learning framework uses only the complementary information to infer the clean signal and ignores the noisy but informative measurements $Y^E$ itself, so it may yield sub-optimal results. The conventional optimization-based denoise methods handle the noisy measurement by exploiting the underlying noise model while also assuming a certain reasonable prior distribution regarding the high-quality signal to regulate the denoised result. Following this denoising philosophy, we also assume that the clean image follows a Gaussian distribution with a position-dependent mean $\mu_X$ and variance $\sigma_X^2$, i.e., $X\sim N(\mu_X,\sigma_X^2)$, which is corrupted by a Gaussian noise \footnote{In practice, the noise distribution in the X-ray CT image is quite complicated and also correlated. To ensure the tractability, in this work, we simply use the independent Gaussian noise as a rough approximation.} with zero mean and position-dependent variance $\sigma_\epsilon^2$, i.e., $\epsilon \sim N(0,\sigma_\epsilon^2)$. Therefore, the noisy image can be modeled as a Gaussian distribution such that $Y \sim N(\mu_X,\sigma_X^2 + \sigma_\epsilon^2)$ if the signal and the noise are independent. Consequently, the conditional probability $P(Y^E |Y^C)$ can be rewritten as:

\begin{equation}
\label{eqn: condition prob}
\begin{array}{lll}
P(Y^E|Y^C) & = & \int P(Y^E|X^E, \sigma_{\epsilon^E}^2) \times P(X^E|\mu_{X^E}, \sigma_{X^E}^2)  \\
     &  &  \times P(\mu_{X^E})dX^E \\
     & = & \frac{1}{\sqrt{2\pi (\sigma_{X^E}^2+\sigma_{\epsilon^E}^2)}}\mathrm{exp}(\frac{-(Y^E-\mu_{X^E})^2}{\sigma_{X^E}^2 +\sigma_{\epsilon^E}^2 })P(\mu_{X^E}).
\end{array}
\end{equation}
where the variances and the mean value can be estimated from the network based on the complementary information, i.e., $\{\mu_{X^E},\sigma_{X^E}^2, \sigma_{\epsilon^E}^2 \}= \Phi_W(Y^C)$. If we substitute equation (\ref{eqn: condition prob}) into equation (\ref{eqn: general self learning}), then perform a negative logarithmic operation, the cost function associated with our PSL framework can be expressed as:
\begin{equation}
\label{eqn: gaussian approx}
\begin{array}{lll}
 \bar{W} & = & \mathrm{arg}\min_W\sum_{t=0}^{T-1}\frac{(Y_t^E-\mu_{X_t^E})^2}{\sigma_{X_t^E}^2 +\sigma_{\epsilon_t^E}^2} + \log(\sigma_{X_t^E}^2 + \sigma_{\epsilon_t^E}^2)  \\
     & &  -\log{P(\mu_{X_t^E})}+ \mathrm{constant}.
\end{array}
\end{equation}

To compensate for the fact that the fluctuation from the noise is stronger than the uncertainty from the image prior distribution, we add a small penalty $-0.1\sigma_{\epsilon_t^E}^2$  to the cost function~(\ref{eqn: gaussian approx}).

Once the network is trained, based on the rule of the Bayesian estimation, the final denoised result of $Y^E$ can be calculated as:
\begin{equation}
\label{eqn: inference}
\bar{Y^E}=\frac{Y^E\sigma_{\epsilon^E}^{-2}+\mu_{X^E} \sigma_{X^E}^{-2}}{\sigma_{\epsilon^E}^{-2}+\sigma_{X^E}^{-2}}.
\end{equation}

From equation (\ref{eqn: inference}), we can see that the variances $\sigma_{\epsilon^E}^2$ and $\sigma_{X^E}^2$ encode the relative importance between the real noisy observation $Y^E$ and the prior knowledge $\mu_{X^E}$.

So far, we have mathematically laid the foundation for both the training and the inference phases, with respect to equation (\ref{eqn: gaussian approx}) and equation~(\ref{eqn: inference}). In the next section, we will detail how to choose the pixels in $Y^E$ and the associated complement $Y^C$, as well as the prior distribution $P(\mu_{X^E})$.

\subsection{Network architecture}
\begin{figure*}[!t]
\centering
\includegraphics[width=0.9\textwidth]{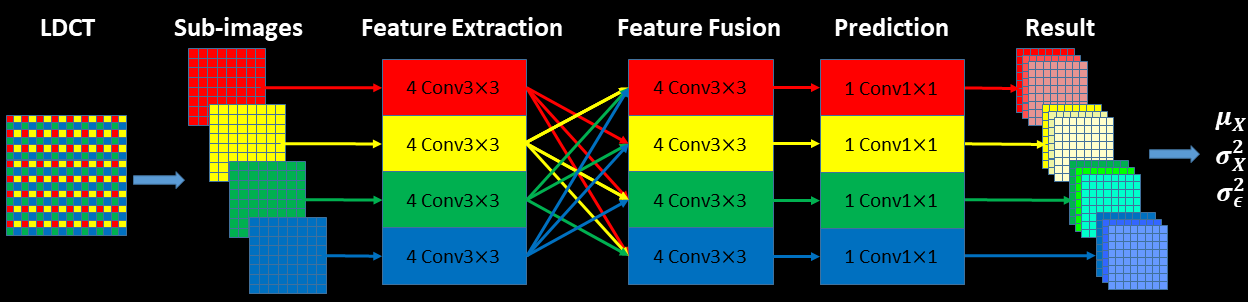}%
\caption{The proposed spatially shift-invariant network architecture. The LDCT is first decomposed into four sub-images, which are fed into the feature extractor. Then, a feature fuser is used to combine the features. Lastly, a predictor is used to generate the results.}
\label{fig: architecture}
\end{figure*}
Because the inherent correlation among pixels plays a vital role in self-learning frameworks, $Y^E$ and $Y^C$ should be highly correlated with each other to ensure the predictability of $\{\mu_{X^E}, \sigma_{X^E}^2, \sigma_{\epsilon^E}^2\}$ given the complementary information $Y^C$. Here, we propose a new architecture based on the shift-invariant property to exploit the inherent correlation of images. The LDCT $Y$ can be down-sampled into four different sub-images, marked as Red for the upper-left ($Y^{UL}$), Yellow for the upper-right ($Y^{UR}$), Green for the lower-left ($Y^{LL}$) and Blue for the lower-right ($Y^{LR}$) sub-images (figure~(\ref{fig: architecture})). In what follows, we will use the abbreviations and the colours interchangeably for convenience, except where it creates ambiguity. 

Obviously, there are strong correlations among these four sub-images since their starting positions are close to each other, for example, by one-pixel shift in our setting. Based on this decomposition, one can randomly choose one of the sub-images as the target pixel set $Y^E$, and let the other three serve as the complementary pixel sets $Y^C$. For example, if $Y^E = Y^{UL}$, the associated complementary information can be set as $Y^C = \{Y^{UR}, Y^{LL}, Y^{LR}\}$, which is fed into the network to calculate $\{\mu_{X^E}, \sigma_{X^E}^2, \sigma_{\epsilon^E}^2\}$. Therefore, for a given image, we have four different combinations, resulting in four different predictions of the mean values, namely $\{\mu_{X^{UL}}, \mu_{X^{UR}}, \mu_{X^{LL}}, \mu_{X^{LR}}\}$. In this paper, we further assume that the mean value, for example $\mu_{X^E}=\mu_{X^{UL}}$, follows a Laplacian distribution with constant variance $\lambda$, whose expectation can be calculated as the averaged complementary mean values $\frac{1}{3}\sum_C\mu_{X^C}=\frac{1}{3}(\mu_{X^{UR}}+\mu_{X^{LL}}+\mu_{X^{
LR}})$. In other words, we have
\begin{equation}
\label{eqn: laplacian}
    P(\mu_{X^E})\propto \mathrm{exp}(-\frac{|\mu_{X^E}-\frac{1}{3}\sum_C\mu_{X^C}|}{\lambda})
\end{equation}

Taking the four different combinations into consideration, omitting the constant term, and substituting equation (\ref{eqn: laplacian}) into equation (\ref{eqn: gaussian approx}), the final cost function can be further instantiated as:
\begin{equation}
\label{eqn: final cost function}
\begin{array}{lll}
 \bar{W} & = & \mathrm{arg}\min_W\sum_{t=0}^{T-1}\sum_{E\in \{UL, UR, LL, LR\}}\frac{(Y_t^E-\mu_{X_t^E})^2}{\sigma_{X_t^E}^2 +\sigma_{\epsilon_t^E}^2} \\
     & &  + \log(\sigma_{X_t^E}^2 + \sigma_{\epsilon_t^E}^2) -0.1\sigma_{\epsilon_t^E}^2  + \frac{|\mu_{X^E}-\frac{1}{3}\sum_C\mu_{X^C}|}{\lambda}
\end{array}
\end{equation}

In this paper, we set $\lambda=1$.

Now, we will elaborate the network architecture $\Phi$, which can calculate the prediction $\{\mu_{X^E}, \sigma_{X^E}^2, \sigma_{\epsilon^E}^2\}=\Phi_W(Y^C)$. It should be noted that one cannot use a single network to deal with different sub-image combinations via parameter sharing, because the position information should be encoded into the network. To solve this problem, we propose a later fusion strategy. The entire network consists of three different modules: a feature extractor module, a feature fuser module, and a predictor module (figure~\ref{fig: architecture}).

Specifically, the input noisy LDCT is first decomposed into four down-sampled sub-images by using the nearest-neighborhood sampling operation with starting positions that have one pixel shifted with respect to the upper-left pixel. The bilinear operation should be avoided here with arbitrary float starting positions, because this would cause information leakage from the target sub-image $Y^E$ to the complementary information $Y^C$, which would lead to a model collapse effect (can only produce $Y^E$ itself without denoising effects).

Then, these four sub-images are fed into four different feature extractor modules. Each module has four layers. For the sake of computational efficiency in this work, we employed the group convolution operator with group number four in the feature extractor module. Thus, the four sub-images input were stacked into a four-channel tensor with a shape of $B\times 4 \times \frac{M}{4} \times \frac{N}{4}$, where $B$ denotes the batch size. The group convolution ensured that the features associated with each sub-image were independent of each other and did not exchange any information.

The complementary information for a target sub-image comes from the other three sub-images. Therefore, a feature fuser module is introduced after the feature extractor module, so that the predictor can make full use of all the information in the complementary pixel sets. To account for the position-dependent property, we use  four different sub-modules in the feature fuser module. Each sub-module contains four layers and accepts only the features from the complementary sub-images. For example, if the target sub-image is the red sub-image in figure~\ref{fig: architecture}, the input to the red feature fuser sub-module will only accept the features from the yellow, green, and blue sub-images.

Finally, a predictor module consisting of a convolution operator and a Softplus activation function $f(x)=\frac{1}{\beta}\log(1+\mathrm{exp}(\beta x))$ is attached to each feature fuser sub-module to predict the associated $\{\mu_{X^E}, \sigma_{X^E}^2, \sigma_{\epsilon^E}^2\}$. In this paper, we set $\beta=1$. We use the Softplus function instead of the more common rectified linear activation function for the following reasons: 1) zero predictions for the noise variance $\sigma_{\epsilon^E}^2$ and the variance of the prior distribution $\sigma_{X^E}^2$ should be avoided, because a zero value would lead to a singularity in the loss function, as shown in equation (\ref{eqn: final cost function}); and 2) the mean value $\mu_{X^E}$ should also be non-negative.

Following the basic design philosophy of modern neural networks, each layer in the feature extractor and feature fuser modules consists of three consecutive operators: convolution, normalization and rectified linear operators. In this work, we adopted the instance normalization operator \citep{RN432} to enable small batch size training. All of the convolution operators in these two modules have a kernel size of $3\times 3$ and a step size of 1. To keep the feature size unchanged, we padded the input of each layer by 1 pixel on each side. Each layer in the feature extractor modules has 64 input and 64 output channels, except the first layer, whose input has four channels because of the number of input sub-images. For the feature fuser module, both the input and output have 192 channels. In the predictor module, the kernel size of the convolution operator is $1\times 1$, and the step size is 1. Since this module will produce the noise variance map, the mean value and the variance maps of the prior distribution for each sub-image, the number of input and output channels associated with each sub-image are 192 and 3, respectively. 
\subsection{Training dataset}
The training dataset we used is a publicly released patient dataset for examining contrast-enhanced abdominal CT images. This dataset was used for the AAPM Low-dose CT Grand Challenge 2016 \citep{RN409}. In this dataset, the original projections were first scanned via routine clinical protocol on Siemens scanners with a reference tube potential of 120KV and reference effective mAs of 200mAs, then reconstructed into CT images by using two different slice thickness, $1\mathrm{mm}$ and $3\mathrm{mm}$. To simulate a low-dose scan, Poisson noise was added to the projection data until it reached a noise level that corresponded to $25\%$ of the normal dose level. The simulated low dose projections were also reconstructed into CT images with $1\mathrm{mm}$ and $3\mathrm{mm}$ slice thicknesses.

In this paper, we used the low-dose CT images reconstructed from the $1\mathrm{mm}$ thick slices, herein termed LDCT, to validate the algorithm’s performance. The normal-dose CT images reconstructed from the $1\mathrm{mm}$ thick slices, herein termed NDCT, served as the gold standard reference. In the officially released dataset, both the LDCT and NDCT are provided in the training dataset, but only the LDCT can be accessed in the testing dataset. Therefore, for quantitative comparison, we split the original 10 training patient cases into a training dataset consisting of eight cases and a testing dataset consisting of the other two. In total, we have 4800 and 1136 2D CT slices for training and quantitative testing, respectively.

\subsection{Testing datasets}
Our testing datasets consist of three different data sources. The first is the above-mentioned new testing dataset derived from the officially released training dataset, denoted as Testing Dataset 1. This dataset contains reference NDCTs and thus can be used for quantitative evaluation.

The second testing dataset comes from the officially released testing dataset of the AAPM Low-Dose CT Grand Challenge 2016, denoted as Testing Dataset 2, which contains low-dose projection data and no NDCT images. Therefore, we converted the original helical-scanned low-dose projection data into axial projection data with 1 mm slice thickness, then used a filtered back projection (FBP) algorithm to reconstruct 2D CT images of size  $512\times 512$, which yielded 445 slices. The FBP reconstruction algorithm uses a ramp filter with a pixel size of $1\times 1 \mathrm{mm}^2$. The major differences between Testing Datasets 1 and 2 lie in their different reconstruction kernels. Testing Dataset 2 is reconstructed with the ramp filter-based homemade FBP algorithm, while the officially released 2D CT images in Testing Dataset 1 is reconstructed with commercial software. Therefore, testing the trained model with Testing Dataset 2 will show the model’s generalizability.

To further test the model’s generalizability with real cases, we acquired Testing Dataset 3 by scanning an anesthetized sheep with lung perfusion at both the low- and normal-dose levels on a Siemens scanner. The exposure settings for the low- and normal-dose levels were 100 kV/150 mAs and 80 kV/17 mAs, respectively. In both settings, we uniformly collected 1160 views over a $360^{\circ}$ range with a $57\mathrm{cm}$ trajectory radius. Each view contained 672 equi-angularly distributed detectors. Both the low- and the normal-dose projections were reconstructed with the FBP algorithm. The reconstruction parameters were consistent with those for Testing Dataset 2. The pixel values were then converted to be CT value and plus 1000 to ensure the consistency with the unit of the training dataset, where the water linear attenuation coefficient was selected to be $0.2\textrm{cm}^{-1}$. It should be noted that the training dataset was collected from patient abdominal CTs, while this real low-dose case was scanned from a sheep lung perfusion study. Therefore, to some extent, this real case can test the out-of-distribution generalizability of the trained model.

\subsection{Training details}
The pixel value of the original CT images is in Hounsfield units, shifted by 1000. As a result, the pixel value of air is 0, while most bone parts have a pixel value around 2000. Therefore, in this work, we normalized the pixel values by dividing by 2000 before feeding the images into the network for training or testing.

To enlarge the training dataset size via data augmentation, we first zero-padded the original image with a size of $512\times 512$ by 16 pixels outside each of the four borders, then performed a random cropping operation to produce another CT image with a size of $512\times 512$. Then, we down-sampled four sub-images with the same image size $256\times 256$ with different starting positions by using the nearest-neighbor interpolation. These sub-images were then fed into the network for training.

We employed the ADAM \citep{RN15} optimizer with hyperparameters $\beta_1=0.9$ and $\beta_2=0.999$ to update the parameters for $1\times 10^{5}$ iterations by minimizing the loss function defined in equation~(\ref{eqn: final cost function}). The initial learning rate was $1\times 10^{-4}$ and diminished by a factor of 10 at iterations of $5\times 10^{4}$ and $7.5\times 10^{4}$. The batch size was 1.

\subsection{Comparison and evaluation metrics}
We compared the proposed PSL method against the well-known total variation (TV) minimization-based denoiser, which needs only the noisy image for denoising because it assumes that the gradient coefficients follow a Laplacian distribution \citep{RN368}. We will also  compare the PSL with the NOISE2VOID (N2V) self-learning algorithm that was proposed to denoise natural images \citep{RN213}, and the CycleGAN method that was a well-known unsupervised learning algorithm.

For the TV method, in this work, we used the Python built-in implementation package ``from skimage.restoration import denoise\_tv\_bregman'', where the hyperparameter is set as 10. This particular strength is chosen based on the visual inspection of image quality that exhibits a good noise-resolution tradeoff. 

In the N2V method, we used the classical U-Net \citep{RN326} as the main network of N2V method. To be specific, since the input has a size of $512\times 512$, to ensure the largest receptive field, we used nine downsampling layers such that the final bottleneck layers have a feature size of $1\times 1$. Besides, doubling strategy is used during the upsampling stage to fuse the features from the encoder and decoder. As for the training part, the input is a LDCT that has been zero-masked on 25 random pixels. The output is the original LDCT. The loss is calculated to be the mean squared error (MSE)-based loss between the predicted 25 pixels and the original pixels. All the other training details were the same as the proposed PSL method.

Regarding the CycleGAN method, a real standard training dataset only requires two pools of datasets from two different domains (for example, LDCT and NDCT in our context) that may not be paired. However, for consistency, we still use the AAPM challenge dataset as the training dataset, which has been used by all the learning-based methods in this work. For a relatively fair comparison, we did not use paired image for training. Indeed, for each training sample, we first randomly choose one image from the LDCT image pool, and then randomly choose one image from the NDCT image pool which is not associated with the previous LDCT image. The code we used for the CycleGAN training is from the official repository (https://github.com/junyanz/pytorch-CycleGAN-and-pix2pix), which has been modified to adapt to our dataset. Specifically, an input image size of $512 \times 512$ is too large to be fitted into the GPU memory for the CycleGAN training. To mitigate this issue, we instead trained CycleGAN by using an input size of $256 \times 256$. In the testing phase, the original LDCT image was first decomposed into four different sub-images by using the nearest interpolation with four different starting point, and then the final processed image was generated by combining the four different sub-images that were processed/denoised by CycleGAN. The generator we used is ``resnet\_9blocks'' which contains nine stacked residual modules. The discriminator we used is a PatchGAN-based classifier which contains three layers. All the other hyperparameters were default. The readers were referred to the paper \citep{zhu2017unpaired} and the code repository for more details.

We first demonstrate the models’ performances on Testing Dataset 1. Because the training dataset was collected from abdominal CT examinations, we selected three different abdominal slices, including one that contains a radiologist-confirmed lesion. Then, we qualitatively compared the denoised images from different methods by visual inspection and quantitatively characterized them with the popular peak signal-noise-ratio (PSNR) and the structure similarity index (SSIM) \citep{RN407} metrics.  Moreover, to evaluate the noise strength, we calculated the standard deviation (STD) of selected flat regions. Besides, we used the contrast-to-noise-ratio (CNR) to quantify the low-contrast lesion detectability. A line profile was also plotted for more detailed comparison.

Then, we compared the denoising performances on Testing Dataset 2. To make a more comprehensive comparison in terms of generalizability, we selected one lung slice, one abdominal slice and one lesion slice. Because there is no NDCT available, we visually compared the denoised images in terms of noise removal and structure preservation abilities.

Finally, we tested the out-of-distribution model performance on the sheep lung data, i.e., Testing Dataset 3. In this case, we also displayed the NDCT for visual inspection, but the misaligned anatomies between NDCT and LDCT prohibit a quantitative comparison. 

\section{Results}
\begin{figure*}[!t]
\centering
\includegraphics[width=0.9\textwidth]{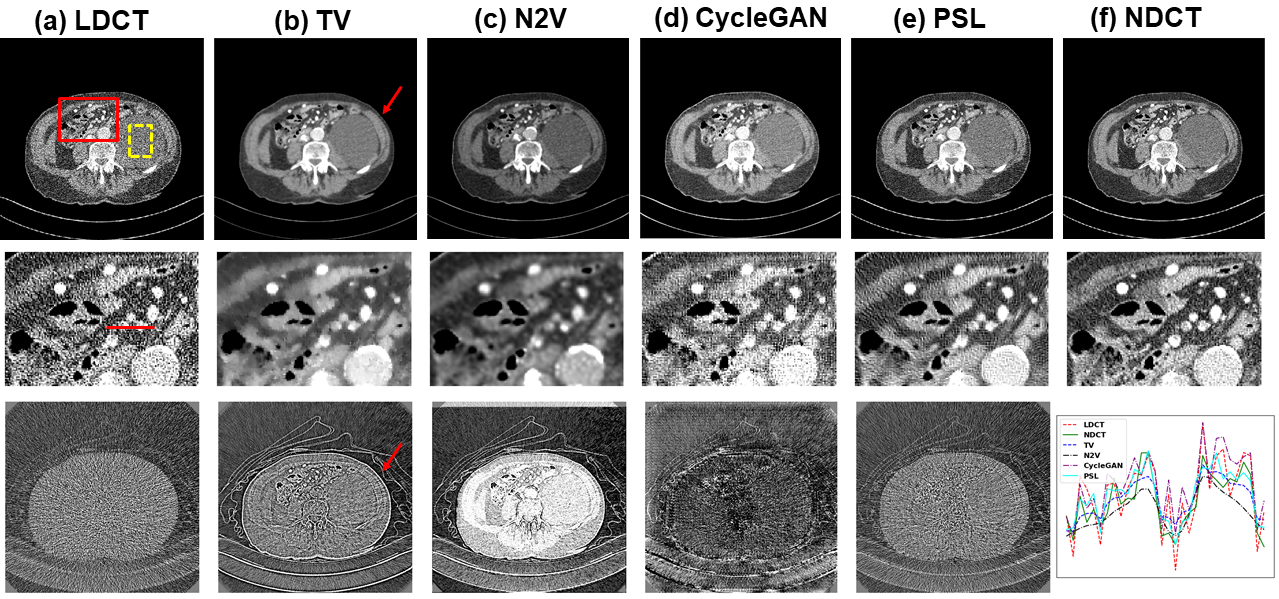}%
\caption{Denoising performance comparison for slice 1 for a patient in Testing Dataset 1. Images from columns (a) to (f) are LDCT, TV, N2V, CycleGAN denoised results, the proposed PSL denoised result and NDCT, respectively. Images from the top row to the bottom row correspond to original images, zoomed-in region of interest, and the difference images against the NDCT image. The display windows for the images in the first and second rows are [-160 240] HU. The display window for the difference images is [-50 50] HU. The lower right sub-figure is the line profile as indicated by the red line in the second row.}
\label{fig: test1 slice1}
\end{figure*}
Figure~(\ref{fig: test1 slice1}) demonstrates the denoised results with different algorithms for a selected slice of a patient in Testing Dataset 1. As we can see in the first row, the LDCT exhibits strong quantum noise, which has been suppressed by the various denoising algorithms. However, both the TV-based and the N2V-based denoisers blurred anatomical structures, which reduces the diagnostic value of their results. In contrast, the proposed PSL method preserves these details better. Compared to the CycleGAN method, the image associated with the proposed PSL method has less noise with comparable, if not better, resolution. The image with respect to the CycleGAN method exhibits brighter intensities, suggesting inaccurate CT values.

These phenomena can be further verified from the zoomed-in views depicted in the second row in figure~(\ref{fig: test1 slice1}). Specifically, many fine structures are overwhelmed by the noise in the LDCT. Although the TV denoiser removed the noise, it yielded the well-known blocky artifacts originating from its piecewise constant assumption. It is not surprising that the N2V denoiser generated overly smoothed details because it considers only the correlation between the target pixels and the neighbouring pixels while ignoring the valuable information from the pixels themselves. The CycleGAN method has noisier and brighter appearance. The proposed PSL method delivers an image with unbiased intensities and more balanced trade-off between noise and resolution.

By further inspecting the line profiles in figure~(\ref{fig: test1 slice1}), we can find that both the TV-based and the N2V-based methods would lead to overly smoothed results, as shown by the dash blue and the dot dash black lines. The strong noise in the result with respect to the CycleGAN-base method (dot dash purple line) can be also clearly observed. By contrast, among all the methods, the proposed PSL method exhibit the smallest difference against the NDCT image.

\begin{figure*}[!t]
\centering
\includegraphics[width=0.9\textwidth]{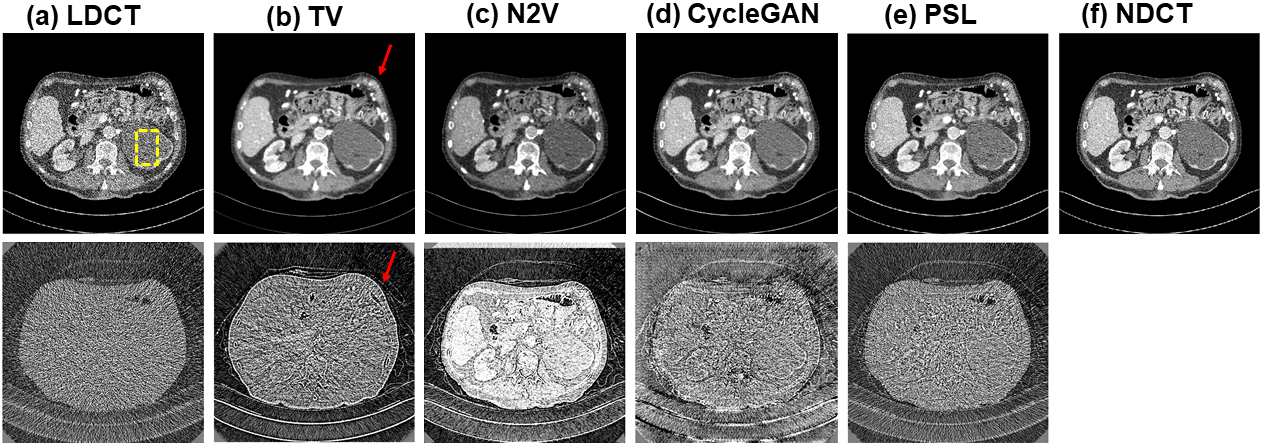}%
\caption{Denoising performance comparison for slice 2 for a patient in Testing Dataset 1. Images from columns (a) to (f) are LDCT, TV, N2V, CycleGAN denoised results, the proposed PSL denoised result and NDCT, respectively. Images from the top row to the bottom row correspond to the original images and the difference images against the NDCT image, respectively. The display windows for the images in the first and second rows are [-160 240] HU and [-50 50] HU, respectively.}
\label{fig: test1 slice2}
\end{figure*}
Figure~(\ref{fig: test1 slice2}) shows another abdominal slice for the same patient from Testing Dataset 1. Again, similar phenomena can be observed regarding the lower noise power in all the denoised images. The proposed PSL method produces higher image resolution than the TV and N2V denoised results, and comparable results as the CycleGAN denoised results. Moreover, by inspecting the body boundaries of the TV denoised images in both figures~(\ref{fig: test1 slice1}) and~(\ref{fig: test1 slice2}), one can find that the CT values are underestimated due to the influence of the air background, as indicated by the red arrows. By contrast, the PSL method retains the skin CT values.

The difference images against the NDCT images depicted in the bottom rows of figures~(\ref{fig: test1 slice1}) and~(\ref{fig: test1 slice2}) show that the TV and N2V denoisers removed many structures, which suggests an over-smoothing effect. By contrast, the difference image with respect to the PSL method is dominated by the noise, and only subtle structures can be observed, which indicates a better image resolution.  The CycleGAN-based method either led to biased CT values (figure~(\ref{fig: test1 slice1})) or removed more structures (figure~(\ref{fig: test1 slice2})). Besides, the N2V method exhibits a severe CT value bias effect towards lower CT values for the whole image, as depicted in figures~(\ref{fig: test1 slice1}) and~(\ref{fig: test1 slice2}).

\begin{table*}
\caption{Quantitative comparison among different methods. Both PSNR and SSIM are calculated based on the whole images, while the STD is calculated based on a small region of interest specified by the dashed yellow rectangles in figures~(\ref{fig: test1 slice1}) and~(\ref{fig: test1 slice2}). CG is short for CycleGAN.}
\centering
\resizebox{\textwidth}{!}{
\begin{tabular}{ccccccccccccccccccc}
\hline
\hline
			& \multicolumn{5}{c}{PSNR(dB)} & & \multicolumn{5}{c}{SSIM} & & \multicolumn{6}{c}{STD(HU)}\\
\cline{2-6} \cline{8-12} \cline{14-19}
 			&LDCT&TV&N2V&CG&PSL& &LDCT&TV&N2V&CG&PSL& &LDCT&TV&N2V&CG&PSL&NDCT \\
\hline
figure~\ref{fig: test1 slice1}&29.04&32.38&30.16&30.25&31.65& &0.6367&0.7570&0.7052&0.7157&0.7201& &109.8&17.11&14.80&70.65&54.30&49.06 \\
figure~\ref{fig: test1 slice2}&26.18&30.59&28.69&29.33&29.34& &0.5511&0.6998&0.6345&0.6828&0.6393& &154.7&34.43&25.10&79.47&66.94&65.50\\
\hline
\hline
\end{tabular}
}
\label{tab: metric table}
\end{table*}
To quantitatively compare the denoising methods, we calculated the PSNR and SSIM values for the different denoised images in figures~(\ref{fig: test1 slice1}) and~(\ref{fig: test1 slice2}) against the NDCT images (Table~\ref{tab: metric table}). Consistent with the visual inspection, all the denoised results improved the PSNR and SSIM values over the corresponding LDCT images. However, the TV denoiser had the best quantitative performance in terms of PSNR and SSIM metrics, despite its visual appearance being inferior to the proposed PSL method. This inconsistency in quantitative and qualitative comparisons has also been observed by other researchers \citep{RN357}. In this paper, we also calculated the standard deviation value (STD) of the selected flat areas (as shown by the dashed yellow rectangles in 
figures~(\ref{fig: test1 slice1}) and~(\ref{fig: test1 slice2})) in different images (Table~\ref{tab: metric table}). The proposed PSL method produced denoised images with STDs comparable to NDCT, but the TV and N2V methods led to smaller STDs, while the CycleGAN method led to larger STDs, which can further justify the over-smoothed (TV/N2V) or over-noisy (CycleGAN) image appearances as observed previously.

\begin{figure*}[!t]
\centering
\includegraphics[width=0.6\textwidth]{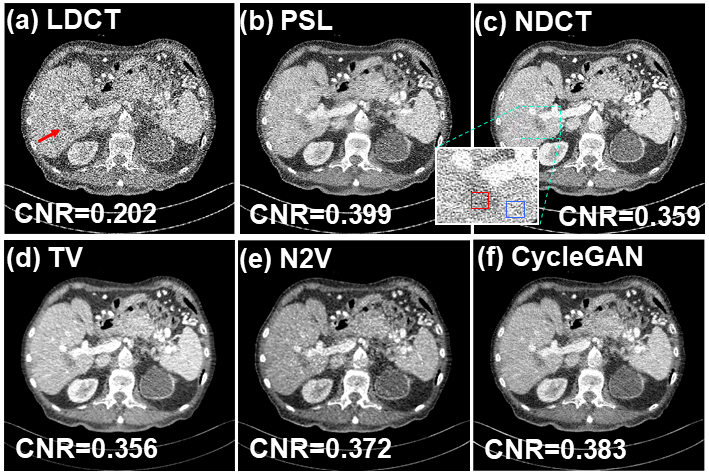}%
\caption{Lesion detectability in slice 3 for a patient in Testing Dataset 1. (a) LDCT, (b) PSL denoised result, (c) NDCT, (d) TV, (e) N2V, and (f) CycleGAN denoised results. The lesion is indicated by the red arrow. The contrast-to-noise-ratio (CNR) was calculated based on the foreground indicated by the red rectangle and the background indicated by the blue rectangle. The display window is [-160 240] HU.}
\label{fig: test1 lesion}
\end{figure*}
Figure~(\ref{fig: test1 lesion}) exhibits the lesion detectability for the images associated with different denoisers, which is more clinically relevant. All of the denoised images from the different algorithms exhibited a comparable detectability for this low-contrast lesion (indicated by the red arrow), despite the PSL method led to a slightly higher CNR value than the others.

\begin{figure*}[!t]
\centering
\includegraphics[width=0.9\textwidth]{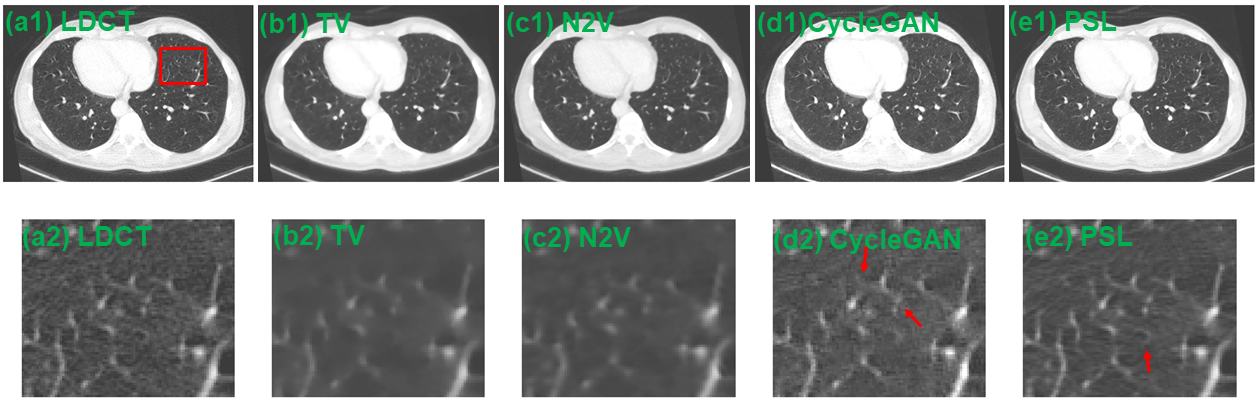}%
\caption{Denoising performance comparison among different algorithms for the lung site in Testing Dataset 2. The images from (a1) to (e1) are LDCT, the TV, N2V, CycleGAN denoised results and the proposed PSL denoised result, respectively. The associated zoomed-in regions of interest specified by the rectangle are shown in (a2) to (e2). The display window is [-1350 150] HU.}
\label{fig: test2 slice1}
\end{figure*}
Figure~(\ref{fig: test2 slice1}) presents the denoised images produced by the different algorithms in the lung area for a patient in Testing Dataset 2. In this dataset, we could only compare the results visually because no NDCT was available. Despite the training dataset only contains a small portion of images belonging to the lung site and also the potential data distribution shift caused by different reconstruction kernels for the training dataset and Testing Dataset 2, the proposed PSL method still achieved a robust denoising performance in terms of preserving the lung nodule details while suppressing the noise. To make it clearer, we also provided the zoomed-in views for the image content indicated by the red rectangle. Clearly, the proposed PSL method delivers an image with less noise than, but comparable resolution to, the LDCT. On the contrary, the TV and N2V denoisers compromise the image resolution, which hampers their diagnostic value. As indicated by the red arrows in Figure~(\ref{fig: test2 slice1}), the CycleGAN method weakened some anatomical structures despite some other structures were strengthened. 

\begin{figure*}[!t]
\centering
\includegraphics[width=0.9\textwidth]{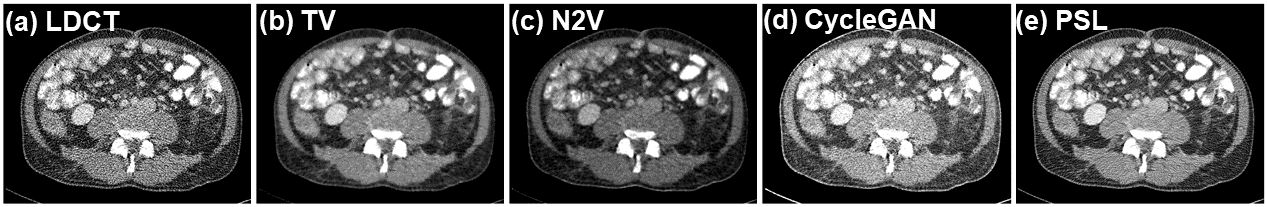}%
\caption{Denoising performance comparison among different algorithms for the abdominal site in Testing Dataset 2. (a) LDCT, (b) the TV denoised result, (c) the N2V denoised result, (d) the CycleGAN denoised result and (e) the proposed PSL denoised result. The display window is [-160 240] HU.}
\label{fig: test2 slice2}
\end{figure*}
Figure~(\ref{fig: test2 slice2}) presents the denoised images for an abdominal slice for the patient in Testing Dataset 2. The image denoised by TV exhibits severe blocky artifacts, and the image denoised by N2V exhibits overly smoothed anatomical structures and a darker appearance. The image generated by the CycleGAN method is relatively noisier and unrealistically brighter. The proposed PSL method yields an image with sharper structures but less image quantum noise, and accurate image intensities.

\begin{figure*}[!t]
\centering
\includegraphics[width=0.9\textwidth]{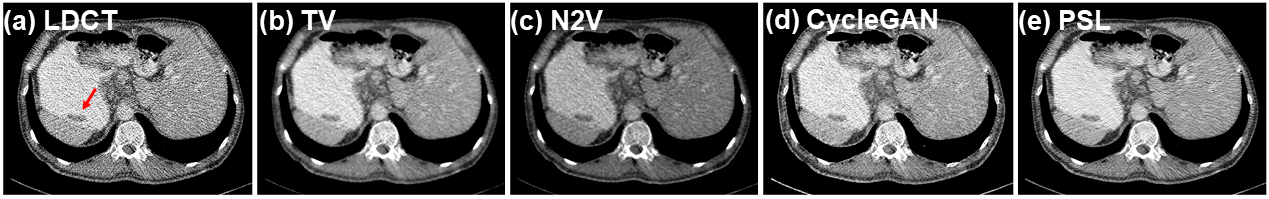}%
\caption{Comparison of different algorithms regarding the lesion detectability in Testing Dataset 2. (a) LDCT, (b) the TV denoised result, (c) the N2V denoised result, (d) the CycleGAN denoised result and (e) the proposed PSL denoised result. The display window is [-160 240] HU.}
\label{fig: test2 lesion}
\end{figure*}
Figure~(\ref{fig: test2 lesion}) compares the lesion detectability of denoised images with different algorithms, indicated by the red arrow. This lesion can be distinguished in all the images, but the edge of the lesion is sharper in the image produced by the proposed PSL method than in those produced by the TV and N2V methods. There remains stronger noise in the image associated with the CycleGAN method than the PSL method. Moreover, the images denoised by the TV and N2V methods exhibit point-like noise despite the structures being blurrier overall.

\begin{figure*}[!t]
\centering
\includegraphics[width=0.9\textwidth]{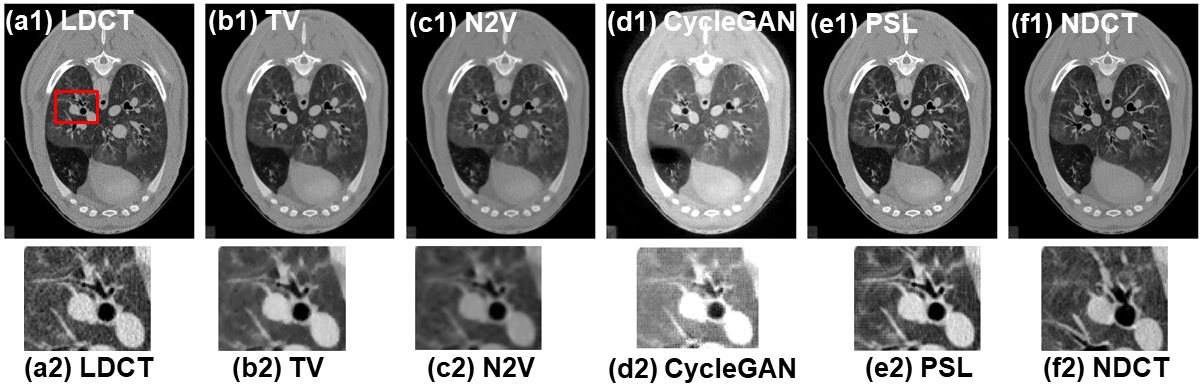}%
\caption{Denoising performance comparison among different algorithms for the sheep perfusion study. The images from (a1) to (f1) are LDCT, the TV denoised result, the N2V denoised result, the CycleGAN denoised result, the proposed PSL result and NDCT, respectively. The associated zoomed-in regions of interest specified by the rectangle are shown in (a2) to (f2). The display window is [-900 900] HU.}
\label{fig: test3 sheep}
\end{figure*}
Figure~(\ref{fig: test3 sheep}) presents the denoised image data from a sheep study with both low- and normal-dose scans. It should be noted that the LDCT and NDCT are not matched exactly because they were scanned at two different time points. Therefore, the NDCT is presented here only for qualitative visual comparison, not quantitative comparison. From the flat region indicated by the red arrow as well as the zoomed-in views in figure~(\ref{fig: test3 sheep}), one can see that the noise has been effectively suppressed by the proposed PSL method. From the internal lung region, one can see that this method has preserved the lung nodule details well. In contrast, the TV and N2V denoisers showed an over-smoothing effect, which diminished their ability to represent small lung nodules distinctly. Another interesting observation is the over high CT values in the image associated with the CycleGAN method.

\section{Discussion and Conclusions}
\label{sec: discussion and conclusion}
Supervised deep learning-based LDCT denoising algorithms require a large amount of pixel-wise paired LDCT and NDCT patient data, which is rarely available in clinical practice. The main purpose of this work is to solve this problem of data scarcity. This issue has never been an obstacle for conventional optimization-based methods, which need only the single LDCT for denoising and which were the mainstream denoising algorithms before the deep learning era. Based on our understanding, the vastly different data requirements for these two groups of methods derive from differences in their driving forces: deep learning-based denoisers are data-driven, but the optimization-based denoisers are model-driven. More specifically, supervised deep learning-based denoisers operate on the hypothesis that the relationship between LDCT and NDCT can be learned by training a neural network on a massive set of pre-collected LDCT/NDCT data pairs. The conventional optimization-based denoisers suppress noise by assuming a prior distribution model regarding the underlying high-quality image, which can characterize the internal pixel correlations, while the resolution is preserved by incorporating a well-explored quantum noise model regarding the measurements. Both the prior distribution model and the noise model are deliberately devised and mathematically described by human experts. Consequently, the denoising performance of these methods highly depends on these mathematical models’ ability to characterize the underlying facts.

Inspired by the strengths of both of these denoising philosophies, we realized that if we could use the powerful deep learning technique to automatically discover the internal pixel correlations and also characterize the stochastics of the noise in a probability viewpoint by exploiting only the LDCT images, we could solve the paired data scarcity problem. To attain this aim, we devised a special convolutional neural network, based on the shift-invariant property of images, such that the statistical property of a part of the LDCT image can be characterized by its associated complementary parts, which then shape our probabilistic self-learning framework. In this paper, we assume that the noise and the underlying clean image follow Gaussian distributions, where the noise is supposed to have a zero expectation. Therefore, the network’s task is to estimate the variances of both Gaussian distributions and the expectation of the image’s distribution.

Extensive experiments demonstrated the superior performance of the proposed PSL method over the TV, N2V and CycleGAN methods. Specifically, despite the TV minimization-based method’s superior ability to remove noise, the associated denoised images illustrate the well-known blocky artifacts, which stem from this method’s piece-wise constant model assumption. Another kind of artifact observed in the images denoised by TV is the underestimated skin CT values, as shown in the TV sub-images in figures~(\ref{fig: test1 slice1}) and~(\ref{fig: test1 slice2}). Currently, the N2V denoising method \citep{RN213} is probably the best known self-learning–based method for natural image denoising. As expected, because it could not incorporate the information from the target pixels, this method led to overly blurred anatomical structures. Additionally, we found that the N2V denoisers yielded images with lower CT values than the other methods. This phenomenon can be clearly observed from the difference between images in figures~(\ref{fig: test1 slice1}) and~(\ref{fig: test1 slice2}). This might be due to the special training strategies. At each iteration, 25 pixels were randomly selected to calculate the gradient for parameter updating. If, occasionally, the majority of these randomly selected pixels belong to the air part (which has a non-trivial chance of occurring, given the large area ratio of air parts to soft tissue parts, as shown in figures~(\ref{fig: test1 slice1}) and~(\ref{fig: test1 slice2})), the network might predict lower CT values to be consistent with the training samples. A more detailed analysis of this benchmark algorithm is beyond the scope of this paper, so we will leave it for future work, for the sake of brevity. It should be noted that this special bias artifact is not observed in the natural image denoising task since there are no such distinctly different pixel values as in the CT images. By contrast, our proposed PSL denoising method exploits the internal pixel correlations by exploring the inherent shift-invariant property, and hence can completely avoid this artifact, which is much more favourable for medical CT image denoising. The CycleGAN method is one of the state-of-the-art methods regarding the unsupervised learning-based image translation task. When apply it on our medical image denoising task, one of the major merits is that it can generate images with similar noise texture as the NDCT images. However, since there is no pixelwise constraint in the CylceGAN framework to enforce image intensity accuracy, the processed image may suffer from inaccurate CT values. This deficiency were clearly observed from figures~(\ref{fig: test1 slice1}),~(\ref{fig: test2 slice2}) and~(\ref{fig: test3 sheep}). It should be noted that in this work, all the results with respect to the compared methods were obtained based on the most standard implementations (for example, based on the official-released codes), their image qualities could possibly be further improved to alleviate the above-mentioned drawbacks. A more comprehensive tweaking for these compared methods is beyond the scope of this work.

Despite its superior performance in enhancing LDCT image quality, the proposed PSL denoiser has several inadequacies that will direct our future research. First, for the sake of easier mathematical description, this paper assumes that image noise follows an independent Gaussian distribution. However, in practice, the noise in CT images are correlated with each other because of the reconstruction algorithms, which results in structured noise artifacts. Unluckily, our shift-invariant–based neural network cannot remove structured noise artifacts because they are also shift-invariant. Actually, these structured noise artifacts can be observed from the zoomed-in view, as shown in figure~(\ref{fig: test1 slice1}). On the other hand, it is well accepted that the noise in the original line integral projection data can be reasonably modelled as independent Gaussian noise. Therefore, a better method might include the projection data during the model training stage. Second, regarding the prior distribution, we hypothesized that the underlying clean image also follows a Gaussian distribution for simplicity, but this will actually be much more complicated in a real-world scenario. Basically, the prior information tries to characterize the distribution of the high-quality clean images. Despite its intractability for an explicitly analytical description, one can train a generative adversarial network (GAN) \citep{RN408} to implicitly describe it given a massive set of high-quality CT images. A higher image quality could be achieved if one can train a denoising network by further incorporating the well-explored noise model of the projection data and utilizing GAN for prior information encoding. We will explore this direction in our future research.

Generalizability remains one of the major concerns during model deployment for all data-driven deep learning algorithms, but especially for medical applications. To examine the generalizability of our trained model, we employed several testing datasets representing different levels of variation, in terms of data distribution, from the training dataset. Testing Dataset 1 was randomly split from the original officially released training dataset (https://www.aapm.org/GrandChallenge/LowDoseCT/), so we assumed that it did not differ from the training dataset distribution. We reconstructed the CT images in Testing Dataset 2 from the projection data in the originally released testing dataset by using our homemade FBP algorithm with a ramp filter, which has a different kernel from the training dataset reconstructed with the commercial software. We expected these different reconstruction kernels to introduce noise distribution shifts in the reconstructed CT images between Testing Dataset 2 and the training dataset, despite the noise distributions being the same in the projection domain. Regarding Testing Dataset 3, which consists of real LDCT/NDCT scans from a sheep study, the noise in the projection domain is supposed to have a different distribution from the training dataset whose noise is numerically simulated, rendering a larger distribution gap than with Testing Dataset 2. Our experimental results suggested a good generalizability of our model among these 3 testing datasets in terms of LDCT image enhancement. However, one should note that many different factors would affect the data distribution, such as different scanners, scanning protocols, anatomical sites, reconstruction algorithms, and noise levels. Therefore, a much more comprehensive study should investigate the model’s generalizability to ensure a robust denoising performance in any real clinical settings. We expect that the self-learning–based denoiser introduced here would achieve better generalizability than the popular supervised learning-based denoiser, since it is easier and cheaper to collect the LDCT images needed to train the former than the paired LDCT/NDCT images required by the latter.

In summary, this paper aimed to solve the problem of paired data scarcity with regard to the deep learning-based LDCT denoising task. To achieve this aim, we proposed a new network architecture to model both the prior image and the noise, which were assumed to follow Gaussian distributions. Specifically, the inherent shift-invariant property was exploited to characterize the pixel correlations. Bayesian inference was used to incorporate the informative measurements. The learning process needs only the LDCT images and thus alleviates the need for paired LDCT/NDCT data. The proposed PSL method increased the average PSNR and SSIM values of LDCT images from 27.61 (PSNR) and 0.5939 (SSIM) to 30.50 (PSNR) and 0.6797 (SSIM). The proposed PSL method also outperformed the TV-, N2V- and CycleGAN-based denoisers, producing an image with well-balanced noise-resolution tradeoff and accurate CT values. Moreover, we examined the performance of the trained model by using datasets with different data distribution shifts, and the PSL showed good generalizability among various testing datasets.

\section*{Acknowledgement}
We would like to thank Varian Medical Systems Inc. for partially supporting this study and Dr. Jonathan Feinberg for editing the manuscript. The public AAPM Low-Dose CT Grand Challenge 2016 dataset used in this study was originally collected from Mayo Clinic under their IRB approval.

\section*{References}
\addcontentsline{toc}{section}{\numberline{}References}
\vspace*{-10mm}

\bibliography{./PSL}
\bibliographystyle{./medphy.bst} 

\end{document}